# SocioEconomicMag Meets a Platform for SES-Diverse College Students: A Case Study

PUJA AGARWAL[1], DIVYA PREM[2], CHRISTOPHER BOGART[2,] ABRAR FALLATAH[1], AILEEN ABRIL CASTRO-GUZMAN[1], PANNAPAT CHANPAISAENG[1], STELLA DOEHRING[1], MARGARET BURNETT[1], ANITA SARMA[1]

[1]Oregon State University, Corvallis Oregon USA

[2]Carnegie Mellon University, Pittsburgh Pennsylvania USA

Emerging research shows that individual differences in how people use technology sometimes cluster by socioeconomic status (SES) and that when technology is not socioeconomically inclusive, low-SES individuals may abandon it. To understand how to improve technology's SES-inclusivity, we present a multi-phase case study on SocioEconomicMag (SESMag), an emerging inspection method for socio+economic inclusivity. In our 16-month case study, a software team developing a learning management platform used SESMag to evaluate and then to improve their platform's SES-inclusivity. The results showed that (1) the practitioners identified SES-inclusivity bugs in 76% of the features they evaluated; (2) these inclusivity bugs actually arise among low-SES college students; and (3) the SESMag process pointed ways toward fixing these bugs. Finally, (4) a user study with SES-diverse college students showed that the platform's SES-inclusivity eradicated 45-54% of the bugs; for some types of bugs, the bug instance eradication rate was 80% or higher.

## 1 INTRODUCTION

Individuals who have low socio-economic status (SES) make up a sizable fraction of the populations that technology needs to serve. In 2020, 37.2 million people in the U.S. were living below the poverty line (an increase of 3.3 million from the year before) [68]. The U.S. Census Bureau estimated that only 50.9% of low-SES households own a desktop or laptop computer, as compared to 96.1% for high-SES [52]. Research has identified challenges that SES-diverse users face [39, 87], finding that individuals from higher SES tend to be more successful in developing career aspirations and are generally better prepared for their careers because of access to resources such as better schools, career offices and guidance counselors, and familial experience with higher education [22]. There is a digital divide in education, where marginal effects of economic, social, and cultural status as a whole impact students' digital skill [18].

In addition to the lack of access to resources, evidence over the past decade shows that technology features also subtly sabotage SES-diverse users' technology usage. Research shows that the way people use software features often differs by SES and, further, many software features are inadvertently designed around the way higher-SES users tend to work with technology [10, 16, 25, 26, 28, 34, 35, 37, 53, 57, 60, 67, 69, 85, 84, 89]. This literature also suggests that, when software features are designed mainly for higher-SES users, it creates barriers for lower-SES individuals. They disproportionately abandon the software when a problem occurs, may refuse to use technology with perceived privacy and security risks, or may accept incorrect outcomes. All of these reactions can impact their ability to secure essential services such as health care, food, housing as well as add to the digital divide in education.

But why would a person's SES be associated with how they use technology? Research suggests different SES levels are characterized by differences in communication literacy, education, culture; in attitudes toward technology risks, privacy, and security; and in perceptions of control and attitude toward authority. We discuss each of these in Section 2.1. These differences affect not only how a person uses software, but also what they need from software.

Consider the example of a hypothetical low-SES student David, who relies on shared devices and works full time at a grocery store in addition to attending a community college. If David is hindered in his use of technology due to the bias built into software he is asked to use in college, it disproportionately disadvantages him. For example, when David encounters technology issues, he may be less likely to request help or extensions from teachers because he believes that the teacher will not act upon his requests—instead he may simply turn in his work as far as he got [60]. In contrast, a higher-SES student might tend to believe that their teacher will accommodate their technological issues and ask for extensions, enabling them to finish their work [60]. In addition, David is also expected to be able to find and evaluate information and assess its reliability for completing his assignments [29]. There is a prevalent digital literacy divide among lower-SES students where they for example might find it difficult to convert files to the correct format and upload assignments to the class Learning Management System [74]. David may neither make equivalent progress to his high-SES peers during class nor catch up on his own because of lack of time and access to technology. These issues are SES-inclusivity issues because they disproportionately impact users like David. Indeed, teachers have reported that low-income students tended to report more obstacles to effective use of educational technology than their peers in more affluent schools [73].

HCI research has produced design recommendations to help low-SES users (e.g., providing positive reinforcement to empower low-income women in using crowd work platforms [82]) or dedicated technology to help low-SES users overcome the digital divide (e.g., Word2Mouth, an eLearning platform to provide educational access for low-income countries [61]).

But, what about existing technology? How can software practitioners even find out if their technology is inclusive to SES-diverse users? SocioEconomicMag (Socioeconomic-Inclusiveness Magnifier, abbreviated SESMag) is an emerging usability inspection method, drawn from extensive foundational research into SES-diverse populations, that enables software practitioners to evaluate problem-solving software for its socioeconomic inclusivity.

This paper investigates how useful the SESMag method is for software practitioners. As an emerging method, it has not been validated yet in the field. In this paper, we present a multi-phase case study to evaluate SESMag's effectiveness in the hands of software practitioners developing real software. We structured our case study into four phases to answer the following research questions:

1. RQ1 (Usefulness): Does SESMag reveal SES-inclusivity bugs, arising from unsupported SESMag facets, in real world-software? (Answered in Phase 1.)
2. RQ2 (Validity): Does SESMag produce valid results? (Answered in Phase 2.)
3. RQ3 (Fixes): How do practitioners use SESMag method to fix SES-inclusivity bugs? (Answered in Phase 3.)
4. RQ4 (Improvement): To what extent did the design changes to fix the bugs make the software more SES-inclusive? (Answered in Phase 4.)

This paper's core contribution is the first evidence of the emerging SESMag method's ability to provide valid, actionable assistance to designers to identify and fix inclusivity bugs that disproportionately affect low-SES users. As such, it makes four component contributions. First, it contributes evidence of SESMag's effectiveness in enabling software practitioners to successfully *identify* SES-inclusivity bugs in their software. Second, it contributes evidence that the SES-inclusivity bugs the practitioners identified using SESMag are *real* inclusivity issues that SES-diverse students actually face. Third, it produces *design remedies* derived by the practitioners from SESMag's facets to fix the inclusivity bugs. Finally, it contributes evidence from a think-aloud study as to the *extent of improvement* the fixes gained.

## 2 BACKGROUND AND RELATED WORK

### 2.1 Background

There is a large body of research investigating how an individual's socioeconomic status (SES) affects their use of technology, through differences in the way they go about reasoning and problem-solving.

*Communication literacy, education, and culture.* Research from multiple countries has shown that an individual's SES is a factor in language literacy, even in their native language [2, 21, 27, 46, 48, 59]. Research further shows that literacy level in the technology's communication language matters to users' success in navigating that technology [16, 35].

Education is a critical factor in an individual's opportunities for literacy-learning activities. However, systemic inequities in education consistently provide education whose quality varies with an individual's and their community's SES stratum, as has been shown in over 40 countries [4, 9, 17, 19, 31, 43, 63, 834]. An individual's educational experience is mediated by the culture with which they identify, and members of minority ethnicities and cultures are disproportionately low-SES; for example, in the US, over twice as many African American and Latino children than White, Asian, and non-Latino children live in poverty [4].

Education about technology is also not equitable across SES levels, such as in providing access to technology-related courses in schools [12, 51, 52]. Due to factors like these, a multitude of significant relationships between a family's SES and their child's technology-related achievements have been shown [1].

*Attitude towards technological risk.* Some demographic groups that are disproportionately in low-SES strata [4] worry more than others about privacy and security risks related to technology usage, and for good reason—such risks materialize for them more often [89, 67]. The steps individuals take to guard against these risks can themselves impact their technology experiences. In addition to using common privacy and security practices like phone locks and passwords, some low-SES individuals go further, avoiding technology risks by avoiding technology altogether whenever possible. [10, 67].

Beyond privacy and security, technological unreliability exposes low-SES users to risk—when using less reliable devices or less stable internet connectivity, technology may fail before an individual can complete the task they are trying to accomplish with it [89, 84]. Even when technological unreliability can be overcome, it may require complex workarounds that make the effort to use the technology outweigh its benefit [25]. Finally, low-SES users risk wasting time without gaining benefit. Families who make ends meet by working multiple jobs [3, 6, 15] often have multiple, competing demands on their time, such as requiring extra time to acquire basic necessities for their children [90].

*Perceived control and attitude toward authority.* Lack of agency and control over one's life are frequently reported feelings by low-SES people [60, 62]. An individual's perception of control over their lives also interacts with their attitudes toward and behaviors with authority figures, and these too vary by SES [30, 36, 78]. Decades of research have strongly correlated low-SES individuals' perceived lack of control with a tendency to be accommodating to authority figures (e.g., [79, 78]).

An individual's perception of control over their outcomes and attitude toward authority directly ties to their behaviors with and around technology. For example, lower-SES students are more likely than higher-SES students to exhibit self-reliance rather than asking for help in the classroom because of negative past experiences of being judged by teachers or peers [60]. Further, some people regard computers as "authorities" or grant computers authority in some circumstances [8, 44, 47, 70]. This research, when coupled with the research on accommodating authority, suggests that individuals' acceptance of technology's errors, dead-ends, or unfavorable results as outcomes to which they have no recourse may vary with socioeconomic status.

User behaviors and perceptions about technology like these that vary across socioeconomic strata, are the focus of SESMag evaluations (explained in Section 3).

## 2.2 Related Work

Over the past 25 years, researchers have investigated ways the socioeconomic digital divide has manifested with ICT (Information and Communication Technology) resources [81]. For example, Dillahunt et al. [24], show that despite the low cost of ridesharing services (e.g., Uber and Lyft), these services are still out of reach for low-income individuals living in low-SES areas. These apps not only have a high digital-literacy requirement, but also presume riders own a smartphone with a data plan, hold a credit card, and know how to initiate rides. Additionally, some drivers equate low-SES areas with safety issues and long commutes, making them reluctant to pick up passengers from these areas. Several other studies have shown that low-SES users face challenges in multiple ways while using ICT devices [24, 75, 76].

The HCI research closest to our work has focused on two categories of how to address inclusivity challenges like the above: (1) by contributing new technologies to support low-SES individuals; (2) by contributing design recommendations to make technology more SES-inclusive; and (3) the InclusiveMag family of approaches analogous to SESMag for other underserved populations.

### 2.2.1 Developing technology for low-SES individuals

In this category, researchers have developed and evaluated technology to help low-SES communities in different fields, including health care [80] and education [61]. For example, in healthcare Saksono et al. developed different apps to reduce obesity in low-SES households. One used "stories" [66] to motivate low-SES users to do physical activity and another used fitness tracking [65] to highlight positive experiences (increase their self-efficacy) and allow reflection on their fitness goals. Other research has focused on low-SES communities that have low health-literacy levels [71] or inadequate access to health care [80]. For example, Khan et al. [45] developed several apps to improve the snacking habits of low-SES families, one of which tries to do so by recommending healthy snacks within a price threshold. Tiwari et al. [77] designed an app to provide maternal health care information to low-literate mothers and provide them with easier access to government health care facilities.

In the education domain, research has investigated ways to remove entry barriers for individuals in low SES situations, including those in low-income countries. For example, Rea et al. designed the Word2Mouth eLearning platform that is designed to work in a slow internet environment and to be easy to use regardless of a user's technology, literacy, and cultural background [61].

Unlike these research projects, our work does not contribute a technology product targeting low-SES individuals. Although it does involve a new technology product, the product is not intended to particularly target low-SES individuals, but rather to be inclusive to individuals across a wide SES range. Our main purpose is to empirically evaluate SESMag's effectiveness, and the product provides an opportunity and platform to carry out that investigation.

### 2.2.2 Design recommendations made for low-SES individuals

In the category of design recommendations for how to adapt technology to address the needs of low-SES individuals, researchers have contributed recommendations relating to low-SES individuals' digital and linguistic literacy [14, 16, 25, 35, 38, 50], as well as perceived control and attitude towards authority [3, 23, 24, 60, 82].

Some design recommendations suggest reducing literacy barriers by using simpler vocabulary [14, 16, 35]. For example, Guo recommended using simplified English when designing technology, such as simple sentence structure, fewer colloquialisms, and fewer references to English culture-specific topics [35]. Similarly, Cheung recommended the use of plain language in professional and technical software interfaces, suggesting that doing so minimizes the cognitive load that low-SES individuals may incur when using these interfaces [16]. Chen et al. also emphasized using familiar metaphors to reduce the technology barrier on social commerce platforms [14], which are heavily used by underserved communities [58].

Several design recommendations focus on addressing the digital literacy divide [25, 38, 50] caused by lack of access to internet and internet-enabled devices among low-SES communities [5, 60]. For example, Marcelino-Jesus et al. recommend providing both online and offline spaces for students to access and share knowledge, since online technology may not always be reliable [50]. Similarly, Hebert et al. recommend making physical (offline) copies of information available [38], to support low-SES individuals who cannot afford the cost of reliable access to Wi-Fi. Dillahunt et al. investigated UX design principles for disadvantaged jobseekers, recommending they be allowed to upload a photographed image of their resume, since lower-SES individuals have limited access to personal computers, but often have mobile phones with cameras [25].

Research also suggests that low-SES students feel less agency and control over their life's outcomes, which translates directly to how and when they use technology [60]. Low-SES students are less likely to believe that their professors will be willing to accommodate technology problems compared to their high-SES peers. Research suggests a few design solutions to overcome such barriers [13, 23, 24, 82]. For example, Beenish et al. describe how mobile apps can be made accessible by following the classic design technique: incorporating back and home buttons in every page to enable users to recover from mistakes or dead-end interactions [13].

Dillahunt et al. and Varanasi et al. talk about the importance of building trust among communities when designing for low-SES populations [24, 82]. Dillahunt et al. suggest that ride-share app developers coordinate with local businesses in low-SES communities to understand how their design can build trust. This will help low-SES users to trust these apps, which they may mistrust due to their low digital literacy and the low transparency such apps often provide about what data they collect and how it is used [24]. Dillahunt et al. also recommend community-based mentoring for individuals experiencing poverty, enabling face-to-face communication among mentors and mentees to build trust among community members. They argue that such technology-mediated mentorship might work best where trust is already embedded within the community [23]. Varanasi et al. talk about building 'technological trust' among communities. In their work highlighting experiences of low-income women on a crowd-sourcing platform, Varanasi et al. observe that women's trust in the app increased when they saw their work being validated, and when consistent payment was reflected in the app [82].

Our work is different from this thread of research, in that we do not contribute design recommendations; we instead evaluate a method for evaluating technology products' inclusiveness across socioeconomic statuses, namely SESMag.

### 2.2.3 Related InclusiveMag work

SESMag was created by following the InclusiveMag meta-method [56]. InclusiveMag enables researchers to construct a new evaluation method for technology inclusiveness, for a diversity dimension of their choice. Here, that dimension is socioeconomic status. InclusiveMag has also been used to create GenderMag [11] to evaluate and improve technology's gender-inclusiveness, AgeMag [55] for age-inclusiveness, and several experimental Mags (e.g., AutismMag, RetinopathyMag, etc. [56]).

To use InclusiveMag on the diversity dimension they have chosen, researchers begin by investigating as thoroughly as possible the reasons why individuals across the chosen diversity dimension might experience technology differently, and group their findings into "facets"—attribute types that can have a wide range of values. For example, GenderMag facets is "technology learning style;" its values range from "tinkering-oriented" (i.e., playfully experimenting to learn a new feature) to "process-oriented" (i.e., wanting to know up-front how the new feature fits into a process) [11].

The facets form the core of an <x>Mag method. Researchers set their <x>Mag facets (and/or personas built around the facet value endpoints) into an analytic process such as a cognitive walkthrough [49]. Practitioners then follow the process to inspect the user-facing portions of their technology through the lens of the <x>Mag facets. For example, Vorvoreanu et al. used GenderMag and its facets to find inclusivity bugs in Microsoft's Academic Search product [86], then derived fixes from those facets. These facet-oriented fixes eliminated the gender gap in performance in their software, in which women had previously experienced twice as many microfailures as men did [86].

Our work has some similarities to the Vorvoreanu study but evaluates the use of SESMag instead of GenderMag. As such, it is the first work to investigate whether the emerging SESMag method can improve technology's socioeconomic inclusiveness.

## 3 THE SOCIOECONOMICMAG METHOD

SocioEconomicMag (Socioeconomic-Inclusiveness Magnifier, abbreviated SESMag) is an emerging usability inspection method for evaluating problem-solving software for its socioeconomic inclusivity. The method aims to enable software practitioners to evaluate whether their software product serves users across the entire socioeconomic spectrum. The goal is to pinpoint any "SES-inclusivity bugs"—ways in which their software may not be socioeconomically inclusive—even if the practitioners doing the evaluation do not have backgrounds in HCI or social sciences.

At SESMag's core is a set of facets of socioeconomic differences that have been extensively investigated in the literature (e.g., [2, 4, 9, 12, 16, 90]). A portion of this foundational literature was summarized in Section 2.1, and Hu et al.'s [41] synthesis of a sizable fraction of it particularly influenced the SESMag facet set. At the outset of our investigation, SESMag had five facets (Table 1). (However, as we explain later in this paper, one of the five facets eventually was split into two separate facets.) As with other members the InclusiveMag family [56], every SESMag facet is regarded as a range of possible values, with an endpoint at each end. For example, the "Access to Reliable Technology" facet ranges from low to high access/reliability

To bring these facets to life, SESMag provides three research-based personas—Dav, Ash, and Fee. As Table 1 shows, each persona has a value for each facet. Dav represents low-SES perspectives, so Dav's facet values are the endpoints that cluster around lower-SES individuals (recall Section 2.1); and Fee's facet values, representing high-SES perspectives, are the high-SES endpoints. Ash represents a fraction of target users with SES between Dav's and Fee's. This endpoint focus is to encourage practitioners who use both Dav and Fee to think about the entire spectrum of facet values. Practitioners who feel that one of the personas is already well-served by their platform can then use only the opposite persona to obtain this cross-spectrum view.

To structure use of these facets and their personas, SESMag embeds them into a systematic process using a specialization of the Cognitive Walkthrough (CW) [49]. The CW is a longstanding analytical method for walking through a use case for a particular technology process from the user's perspective. The SESMag specialization explicitly embeds a SESMag persona as well as the facets into this walkthrough. For example, Figure 1 summarizes how SESMag works with the SESMag-specialized CW and a persona named "Dav" who represents facet values common among low-SES users.

To find the SES-inclusivity bugs, evaluators walk through the use-case. They begin by choosing the first/next in-the-head subgoal that they *hope* a user has, then evaluate whether the selected persona *will* have that subgoal, using the persona's facet values to reason it out. They then choose actions (in-the-fingers) they hope a user will perform to carry out that subgoal, decide whether their selected persona will indeed do those actions and, after the action, if their selected persona will see that they're making progress towards their goal. The evaluators continue looping through subgoals and action until the end of the scenario.

Table 1: SESMag method's five facets, and Dav's, Fee's, and Ash's facet values. See also text in Section 2.1

| SESMag facet, value range | Facet value for each SESMag persona |
|---|---|
| **Access to Reliable Technology:** Access to reliable devices with reliable internet connection. Includes owning vs. sharing vs. depending on public devices. *Facet's value range:* Low to high access/reliability | *Dav*: seldom has access to reliable tech device(s)/software & internet. E.g., they may only have an older phone for their tech tasks [34] <br> *Fee*: Usually has access to reliable, up-to-date tech devices/software & internet. <br> *Ash*: Situational. Closer to Dav in some (e.g., at home), closer to Fee in some (e.g., at school), and between Dav & Fee in others (e.g., when with friends). |
| **Communication Literacy/Education/Culture:** Fluency of understanding & communicating in the tech's target language (e.g., vocabulary, idioms, cultural references) and their quality/level of education. *Facet's value range:* Low to high alignment of communication literacy/education/culture compared to that assumed by the technology | *Dav*: Low alignment. E.g., educational background provided different cultural references, literacy, or tech background than those upon which the technology relies. [43, 27] <br> *Fee*: High alignment. <br> *Ash*: Between Dav & Fee. |
| **Technology Self-Efficacy:** Self-perceived ability to use technology and perform computer-based tasks. *Facet's value range:* Low to high self-efficacy | *Dav*: Lower self-efficacy than their peers about unfamiliar tech tasks. May underestimate their ability to succeed, maybe quit if problems arise. E.g, Dav attended under-resourced schools with little tech education [43] so does not view themself as being "good at tech" [42]. <br> *Fee*: Higher self-efficacy than their peers with technology. <br> *Ash*: Medium self-efficacy with technology. |
| **Attitudes toward Technology Risks, Privacy, and Security:** Willingness to take technological risks, ranging from using unfamiliar features to taking risks with privacy/device security. *Facet's value range:* Low to high tolerance for technological risks. | *Dav*: Low tolerance for tech risks. E.g., Dav may mistrust financial apps due to potential questionable usage of their financial information or risk of personal identity theft [84]. <br> *Fee*: Higher tolerance for tech risks. E.g., Fee is careful with passwords, but is willing to use apps that collect personal and financial information. <br> *Ash*: Between Dav & Fee. |
| **Perceived Control and Attitude toward Authority:** Perception of their ability to exert influence over tech's positive or negative outcomes and interactions. *Facet's value range:* Low to high perceived control | *Dav*: Low perception of control over tech outcomes, especially when tech outcomes are extensions of authority figures. E.g., Dav is unwilling to turn to authority figures for assistance when facing tech difficulties [33, 60]. <br> *Fee*: Higher perception of control over tech outcomes. E.g., Fee is willing to explore workarounds or personal follow-ups. <br> *Ash*: Between Dav & Fee. |

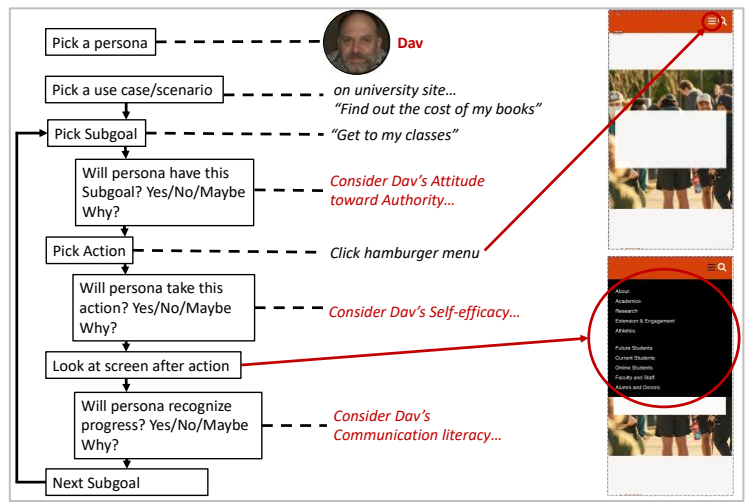

Figure 1: A practitioner's workflow to use SESMag method, showing how the evaluation loops through the sub-goal and action questions in the method (left), along with user-interface components being evaluated (right).

## 4 METHODOLOGY

To investigate the SESMag method, we conducted a case study. A case study is an empirical investigation, drawing from multiple forms of evidence, into a specific phenomenon in a specific real-life context [64]. Our "case" was the use of SESMag by a real team engaged in a real software development project. The context was Team C, the learning platform they were developing, and its evaluation. The case study ran for 16 months and followed Team C in their process of developing a learning management platform for college students (including underrepresented and low-SES students).

The study included four phases, one for each RQ. For RQ1, in Phase 1, Team C conducted an SESMag evaluation on their prototype to identify SES-inclusivity bugs that would disproportionately affect lower-SES individuals. For RQ2, in Phase 2, Team C conducted online interviews with a faculty member and a student at a regional college where 70% of undergraduate students are on financial aid, to validate the SES-inclusivity bugs they had found in Phase 1. For RQ3, in Phase 3, Team C made design changes on their digital prototype for the validated bugs. For RQ4, in Phase 4, Team C conducted a controlled think-aloud study with seven participants in their target population to test the efficacy of the fixes found.

We used triangulation to safeguard rigor. Triangulation [20] draws on convergent evidence from multiple sources to strengthen the validity of a claim. We used three triangulation mechanisms—data triangulation, triangulation with research literature, and investigator triangulation. We used data triangulation extensively, triangulating what Phase 1 evaluators wrote down with video recordings of what they said, and triangulating Phase 2's interview with the instructor with that of the student interviewee. Phase 1 also used triangulation with research literature. In Phase 2 and Phase 4, we also used investigator triangulation to safeguard the rigor of the qualitative coding efforts using independent coding by different researchers and then calculating inter-rater reliability.

### 4.1 Phase 1: Team C uses SESMag to evaluate their platform

When our case study began, Team C was in the process of building a digital prototype of an online learning platform for students taking technology courses at community colleges and 2-year or 4-year universities. Team C had prepared the prototype of the platform on Figma, a web application for design. Team C was interested in SESMag because their platform was part of Team C's vision of creating pathways to technical employment for underrepresented and low-SES students.

The members of Team C who participated directly in the case study were a research software engineer (C-1), a user experience engineer (C-2) (neither of whom had prior experience with SESMag), a product design intern (C-3), and a platform support technician (C-4). These team members' actions were also influenced by the other members of Team C, whose members included a project lead who maintained the team's vision, a design team who iterated on the user interface, an infrastructure team of 3 engineers who will bring the design prototypes to production, a content authoring team who owned the curricula presented on the platform, and an advisory group of college instructors and educational researchers.

In Phase 1, Team C used SESMag to evaluate their current prototype. To prepare for the upcoming evaluations, C-2 first met with Researchers R-1 and R-2 to choose the scenarios to evaluate (Table 2) and which SESMag persona(s) Team C would use. Because of their interest in low-SES and underrepresented students, Team C chose the low-SES "Dav" persona for the evaluations, whose facet values were described in Table 1. Recall that Dav represents endpoints of the facet values that align with lower-SES perspectives. Team C already had experience with higher-SES students, and so their intent was to emphasize facet values of users that Team C was not already familiar with. Portions of the SESMag personas that are not about the facets (e.g., appearance, demographics, experience, job title, etc.) are customizable, so Team C customized their Dav persona to be a community college student who also has a full-time job working at a grocery store (Figure 2).

Table 2: Team C decided to evaluate four scenarios. *(Scenario 2 originally said "project submission", but was evaluated as a "task submission", which is subtly different in a way that turned out to be important.)

| Scenarios |
|---|
| 1. Student wants to start the first assignment due this week. |
| 2. After a \<task\>* submission, student wants to check their submission status and what score they got. |
| 3. Halfway into the semester, student wants to check if they will pass this course. |
| 4. Have I submitted my assignment due in 2 days? |

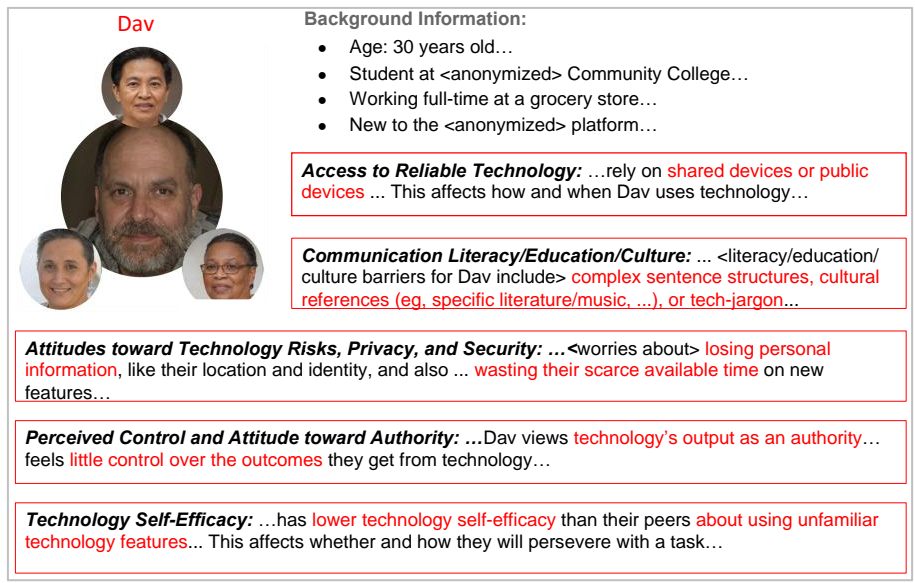

Figure 2: An excerpt of low-SES Dav persona that Team C customized. The top (background info) is customizable: they made Dav a community college student also working full-time at a grocery store. The rest (5 red boxes with the facet values) is research-based, so not customizable. The full persona is in the supplemental document.

Team C evaluated the four scenarios via two 2-hour SESMag sessions (Zoom, with videorecording). Present at the sessions were C-1, C-2, C-3, C-4, and members of the research team who helped to coach and make observation notes. Team C noted their step-by-step evaluation decisions on the SESMag forms (supplemental document), which then served as their SES-inclusivity "bug report". At the end of each session, Team C did a debriefing to review the bug report they had just created, to discuss suggestions for solutions, and to plan for follow-up interviews with community college members to assess the validity of what they had found. Additionally, the researchers reviewed the session notes, the SESMag forms, and the meeting transcriptions to clearly map the facets to the bugs, and removed one facet (for Bug3) that was not justified in the evaluation session.

Using lessons learned from Team C's experiences, we made a few improvements to the SESMag method. In Team C's first session, they used the five facets Hu et al. had proposed [41], which combine the Risks and Privacy/Security facets in Figure 2 into just one Risk facet. During the evaluation, we observed that this combination seemed too general, with evaluators overfocusing on some aspects of risk (lack of time) to the extent that they overlooked other aspects of risk (privacy). Thus, before the second session, we split that facet into two, using a total of 6 facets in the second CW evaluation session. To maintain consistency for analysis, one member of research team qualitatively coded the first session's forms data using Table 3's codeset (described next, in the Phase 2 section) to split the overly large facet "Attitudes towards Technology Risks, Privacy, Security" facet responses into Privacy/Security vs. other Risks.

## 4.2 Phase 2: Team C interviews stakeholders to validate SES-inclusivity bugs

In Phase 2, Team C interviewed an instructor (I-1) and a recent low-SES college student (S-1) from one of the 12 colleges who would eventually be using the platform; the college has a high proportion of low-SES students. The goal was to find out whether issues like the SES-inclusivity bugs Team C had found during Phase 1 had ever arisen among CS students at their particular college.

Team C conducted the interviews, which were semi-structured, one at a time over Zoom. The interviews were video-recorded, and later transcribed. Researchers were also present, making observation notes along the way and occasionally contributing to the questioning.

In the interview, for each bug Team C had noted on their SES-inclusivity bug report during Phase 1, Team C explained the scenario and showed the prototype at the point the bug had arisen, and then asked the interviewee whether such a bug had arisen among their college's CS students. The interviewees were asked to make this assessment based on whether they had previously *experienced* or *witnessed* that kind of bug among these students.

R-1 to S-1: *"...we want you to think of yourself as a historian and/or an anthropologist where you're telling us, particularly from the viewpoint of you and your friends...".*

Team C showed the prototype up to the point of the bug in question, and explained what a student was supposed to have done up to that point in time, and then asked:

C-2 to S-1: *"The next action...is to read this submission table and identify his grade. Is this what you would do?"*

S-1: *"Yes. I would be confused as to how to read it, though..."*

C-2 would then follow up on the interviewee's response by asking as many questions as needed about why they had answered the way they did.

At the end of the interview, Team C also asked a final set of (unstructured) follow-up questions intended to elicit discussion of characteristics that correspond to facet values the interviewee may have seen among students at their college, such as the equipment they used to complete their assignments (relating to the technology access facet).

Using content analysis [40], two HCI researchers independently coded 30% of the instructor interview data using the codeset in Table 3, achieving an inter-rater reliability rate (IRR) of 95% (Jaccard) [72]. They also used the same codeset to independently code 20% of the student interview data, achieving an IRR of 99% (Jaccard). Given these high IRRs, one researcher then finished up the rest of the coding.

Table 3: Bug presence/absence codes, and Facet codes, used to analyze Phase 2 interviews. (Underlines show abbreviations.)

| Code and Definition | Excerpts from interviews |
|---|---|
| **Bug Presence Confirmation Codes** | |
| <u>Yes</u>: Respondent has witnessed that bug in the students at their college. | S-1: "...From my last computer class they were all PDF templates and that conversion from word from PDF to word was sometimes complex ... so that was a big issue ..." |
| <u>No</u>: Respondent has had the opportunity to see the bug (i.e., witnessed the scenario in which the bug would arise), but the bug did not arise. | I-1: "...they probably try to click on the weekly activities. They'd probably try to search out more information right there..." |
| <u>N/A</u>: Respondent has had no opportunity to see the bug arise (i.e., never witnessed the scenario in which the bug would arise.) | I-1: "... they won't reach out ... so there is an issue or there is a concern we won't know about it." |
| **Facet Codes** | |
| <u>Access</u> to Reliable Technology: Respondent mentions students' access to reliable devices or reliable internet or reliability on shared devices or public devices. | S-1: "wi-fi was inconsistent because it's at the school campus and so sometimes it's really good and then sometimes it really sucks; it really depends on the floor, you are on actually." |
| <u>Comm</u>unication/Educ/Culture: Respondent mentions students struggling with complex sentence structures; cultural references or tech-jargon; education quality or education compared to peers involving up-to-date technology; and/or frequency of reading complex text. | I-1: " there's a lot of our students that come in, especially in the CS side that have no prior technology experience." |
| Attitudes toward Technology <u>Risks</u>: Respondent mentions students' attitude toward risk in using unfamiliar technologies that they might need to spend extra time on, using familiar features to perform tasks. | I-1: "..one of the big things about our students is there's so many working adults in our program. And so many adults with children and especially young children, just it's hard for them to find time... and I worry about the related primers." |
| Technology <u>Priv</u>acy and Security: | <No mentions of this facet> |
| Perceived <u>Control</u>, Attitude to <u>Authority</u>: Respondent mentions students viewing technology's output as an authority, and/or lack of control over the outcomes from technology. | I-1: "it's a matter of I'm paying for this already I'm paying for tuition and now you're telling me, I have to pay something else...And the veterans ... whatever the GI bill pays for, that's <all they can afford>" |
| Technology Self-<u>Eff</u>icacy: Respondent mentions students' technology self-efficacy/confidence about using unfamiliar technology features, including whether they blamed themselves (instead of the tech) when problems arose with technology. | S-1: "if it was MAC computer I couldn't install half the computer science tools, I needed and then same thing applies to people who are using Linux distributions.." |
| <None>: Respondent did not mention any of the SESMag facets. | S-1: "... the title it doesn't associate with the project name." |

### 4.3 Phase 3: Team C uses Facets to Find Fixes

In Phase 3, Team C then fixed many of the SES-inclusivity bugs that had been validated by the Phase 2 interviewees. Toward this end, Team C turned again to the bug report that they had created during Phase 1.

The bug report elicited not only SES-inclusivity bugs, but also *why* the bug might arise—what specific problem-solving facet(s) were not supported in the design of the interface. Team C used these facets identified in Phase 1 as a starting point for fixing the bugs. Once Team C decided on design changes, they created a revised version of the prototype. For the rest of this paper, we refer to the prototype that existed at the beginning of Phase 1 as the "Original" version and the prototype that Team C produced during Phase 3 as the "post-SESMag" version.

### 4.4 Phase 4: Team C evaluates their fixes

In Phase 4, to find out whether the fixes from Phase 3 were effective at removing the SES-inclusivity bugs from Phases 1 and 2, Team C performed a think-aloud study with seven SES-diverse students, using the post-SESMag prototype.

#### 4.4.1 Procedures

Team C conducted the think-aloud user study with one participant at a time over Zoom. Team C's member C-2 was the study facilitator and ran each session with the participants. Each session lasted up to 70 minutes and was video recorded.

Each session began with the participant reading and agreeing to the IRB consent form, then filling out two Likert-style questionnaires (with a few other types of questions) about their SES identity and SESMag facet values, respectively, and finally doing a think-aloud warm-up/practice. They were instructed to: "... thinkaloud as you complete a few scenarios through a prototype of a learning platform tool. What I mean by thinkaloud is that you will verbalize all your thoughts as you interact with the tool." (Full script is in the supplemental document.) The participant then began the main task.

For the main task, each participant began using Team C's post-SESMag prototype. The post-SESMag prototype was a Figma prototype, which the Team C facilitator shared by providing the participant with its URL using the Zoom chat. The participant opened the URL on their own machine with Zoom screen-sharing on.

They used this prototype to work through Scenarios 1, 2, and 3 from Section 4.1's Table 2, with context wording added, such as "You're in the Cloud Admin online course. Start and read…" When a participant finished all three scenarios, we compensated them with a $20 Amazon gift card. The questionnaires and scenarios we used are in the supplemental document.

#### 4.4.2 Participants

To recruit both lower- and higher-SES participants, Team C recruited 1st-year CS students from a U.S. community college and from a U.S. public university. The community college was located in a state whose mean household income matches the national average; the U.S. public university was in another state whose mean household income is in the bottom 25% [88, 12]. Participants were recruited by email, by word of mouth, and by asking instructors to distribute a recruitment flyer to their classes. We selected the first seven participants who matched our criteria. All participants were at least 18 years of age.

We did not collect traditional demographics beyond being over 18, but we did collect socioeconomic identity (SES identity) via one of the questionnaires. That questionnaire contained nine, 9-point Likert-style questions about their socio+economic identity: their perception of their food security, of their minority status, of their parents' education compared to that of others in their state, etc. (see supplemental document for full questionnaire). One participant did not fill out the SES identity questionnaire. Of the other six participants, three identified as higher SES (more SES-identity responses above "neutral SES" than below) and three identified as lower SES (more SES-identity responses *below* neutral).

#### 4.4.3 Analysis

Two HCI researchers independently used content analysis [40] to code participants' video-recorded actions and verbalizations, as to whether participants had still encountered the bugs from Phases 1 and 2, despite the fixes from Phase 3. Specifically, two researchers independently coded 19% of the data using the code set in Table 4. The researchers' IRR was 89% (Jaccard). Given this high agreement, one of the researchers completed the rest of the coding.

Table 4: Codes for whether a Phase 4 participant ran into one of the inclusivity bugs that Team C had attempted to fix in Phase 3.

| Code, Definition |
| --- |
| Fixed ($\sqrt{}$): The participant had an opportunity to experience the SES-inclusivity bug—i.e., got to the same situation in which it had arisen in Phases 1 and 2—but did not experience the bug. |
| Fixed-ish ($\sqrt{}$?): As with "Fixed", the participant with an opportunity did not experience the bug—but they did not seem to entirely understand it either. |
| Not Fixed (X): The solution provided for the SES-inclusivity bug did not fix it: the participant experienced it despite the design change. |
| N/A: The participant did not have an opportunity to experience the bug (e.g., did not navigate to the place where the SES-inclusivity bug from Phases 1-2 could have arisen). |

## 5 RESULTS

### 5.1 RQ1 Results (Usefulness): Team C uses SESMag

Since the purpose of the SESMag method is to find SES-inclusivity bugs in software platforms, we begin by answering the first research question: Does SESMag reveal SES-inclusivity bugs, arising from unsupported SESMag facets, in real world-software?

Team C found SES-inclusivity bugs in 76% of the questions answered in the two SESMag evaluation sessions. To calculate this fraction, we defined a bug as being a SES-inclusiveness bug if the team used one or more of the facets in their persona to identify it, because the facets (attitude toward risks, privacy & security, etc.) represent the empirical findings of individual differences clustered by socioeconomic status. The first walkthrough covered seven questions about Dav's subgoals and actions. For all seven questions, the group (Team C and members of research team) identified inclusivity bugs they decided Dav might face, and in each case the group associated these bugs with the facet value(s) that explain why the inclusivity bug would arise. The second walkthrough covered ten questions; six of these revealed bugs, all of which they explained with Dav's facet values (Table 5).

Table 5: Team C's SESMag evaluation summary. 76% of questions included bugs, all of which were SES-inclusivity bugs.

| Team C | # questions | # bugs revealed | # SES- inclusivity bugs revealed | % questions revealing SES-inclusivity bugs |
| --- | --- | --- | --- | --- |
| Session 1 | 7 | 7 | 7 | 76% |
| Session 2 | 10 | 6 | 6 | |

To validate that the group had been accurate in writing down their analysis, we triangulated their bugs and facet reasoning against the video recording. All but one facet assignment triangulated; we excluded that case because there was no evidence it wasn't inadvertent: it had no explanation in writing and no mention in the video. Table 6 summarizes the final set of SES-inclusivity bugs Team C found in Phase 1 and the facets identified.

These facets played an important role in focusing Team C's attention during the evaluations. Facets are the SESMag method's primary technology-transfer elements: they make the research findings actionable about characteristics common in low-SES users. For example, one of the questions Team C evaluated was "Dav wants to check if they have submitted their assignment due in two days". This required a series of steps, including first scanning a group of upcoming assignments in the homepage to see which assignments were due. When evaluating the question "Will Dav know they did the right thing after scanning the homepage's table" (Bug13), the evaluators identified an SES-inclusivity bug. They decided Dav would be overwhelmed seeing that multiple different assignments were due in 2 days and not understanding which of them were a priority (Risks facet), as the page didn't mention which assignments were graded and instead relied on the students' ability to differentiate between the concepts of primer and project (Comm facet).

Using the Dav persona allowed Team C to identify a set of potential SES-inclusivity bugs (RQ1). However, aware that they as software developers were not Dav-like users themselves, the team decided to validate the identified SES-inclusivity bugs, through interviews with an instructor and a potential user.

Table 6: SES-inclusivity bugs, facets, and triangulation. Columns 1-2: Phase 1 results. Column 3: triangulation with research literature. Column 4: Phase 2 data. All Phase 1 results were consistent with research and with problems witnessed/experienced by the Instructor (I) and/or Student (S).

| Phase 1 data: | | Consistency with: | |
|---|---|---|---|
| **SES-inclusivity bugs: Someone like Dav...** | **Facet values (Table 3 abbreviations) + Why (excerpted from evaluation forms)** | **Research Literature relating to excerpt** | **I/S** |
| *Bug1*: Might not start the first assignment due this week: Big Data Analytics. | Risk: "Dav doesn't have <that much> time… grocery store shift starts in an hour…" Other facets used: SE | Risk includes aversion to <u>wasting time</u> on tech dead-ends [3, 6, 15] | I |
| *Bug2*: Might not click the "ongoing" button for Big Data Analytics. | Access: "...using an <u>ancient laptop</u>… flaky keyboard... may actually be pounding on <unclickable> Big Data Analytics" Other facets used: SE | Access: seldom access to <u>reliable device</u>(s)/software & internet [34] | I S |
| *Bug3*: After clicking "ongoing" button, might not know was right/see progress. | SE: "I don't see the word "assignments", so <u>I'd be doubtful</u> this was right." Other facets used: Risk, Comm, Ctrl/Auth | SE: lower tech self-efficacy including <u>underestimating ability</u> to succeed [42] | I S |
| *Bug4*: Might not read through the page and click Next. | Comm: "maybe <u>scared by techy words</u> he doesn't know: MapReduce, Parallel Processing..." Other facets used: SE, Risk | Comm: understanding the <u>wording</u> (including jargon) & education [27, 43] | I S |
| ... (*Bug5-Bug7 in Appendix A*) | | | |
| *Bug8*: Might not check ... after a project submission... what score they got. | Privacy: "It is not clear for dav <that grades tab makes sense here>... so… <u>risk to privacy and security.</u>" Other facets used: Risk, Ctrl/Auth | Privacy: includes <u>distrust of losing personal info</u> to unexpected features [33,60] | I S |
| *Bug9*: Might not click "grades" near the upper left corner. | Ctrl/Auth: "There's <u>no indication</u> that he made a submission..." Other facets used: Risk | Ctrl/Auth: includes <u>low perception of control over tech outcomes</u> [33, 60] | I S |
| ... (*Bug10-Bug13 in Appendix A*) | | | |

## 5.2 RQ2 Results (Validity): A College Instructor and Student Weigh In

To analyze whether the SES-inclusivity bugs Team C had found were real issues for the low-SES users in the target audience (RQ2), Team C interviewed a college instructor and student (as detailed in Section 4.2). The results showed a very high validity rate: the instructor confirmed that they had seen their students having related problems with all except one of the bugs; the college student also validated all except one of the bugs, using their own experience and those they had seen in their peers' experience. Together, their confirmations validated 100% of the SES-inclusivity bugs Team C had identified in Phase 1 (rightmost column of Table 6).

In explaining why and how they had seen bugs like those in Table 6 arise, the interviewees often referred to concepts closely related to Dav's facet values. For example, when reviewing Bug13, they referred to usage characteristics identified by Risk and Comm facets.

I-1: *"they would probably look at it [table], all, I have to do all of them[assignments]… If didn't understand the difference between the primer and the project."*

S-1: *"I agree… if I had not gone through the introduction yet, I just assume they're all the same priority and they all need to be done, like ASAP."*

S-1: *"If I did not know what the difference was between primer and project, I would assume all three were [required] but if I did know yeah, I'd make the assumption that project is what's due in two days."*

In fact, qualitative coding (recall Table 3) showed that the instructor and student referred to these concepts more than 50 times. Although the instructor and the student had not been introduced to SESMag before (they did not know SESMag's facets or facet values), a lot of their explanations during the unstructured part of the interview helped confirm the pertinence of Dav's facet values:

I-1 *(Education aspect of Communication Literacy/Education/Culture facet): "yeah there's a lot of our students that come in, especially in the CS side that have no prior technology experience...."*

I-1 *(Communication and culture aspects of Communication Literacy/Education/Culture facet): "You know because you have in here a whole host of words ... they would want to say, 'Well, you know, what's the example? How's this [similar to] what have I done in my life?…"*

S-1 *(Risk facet's aspect of having little spare time to waste): "... I was working two jobs, so I only had time to do it probably like right before the due date..."*

I-1 *(Self-efficacy facet's aversion to unfamiliar tech): "... I'm not sure that they would feel 100% comfortable with <document database system> so they probably jump into <alternative>."*

Now in possession of a list of SES-inclusivity bugs with the facets to explain their causes (Phase 1), along with the knowledge that the bugs really occur (Phase 2), Team C had enough information to redesign the platform to fix them.

### 5.3 RQ3 (Fixes) and RQ4 (think-aloud) Results: Team C Redesigns and Evaluates their Platform

How did Team C fix these bugs (RQ3), and how well did their fixes work for the SES-diverse college students in our study (RQ4)? In this section we present the results of Phase 3 (how Team C fixed the inclusivity bugs) and Phase 4 (the think-aloud study with SES-diverse college students), which answer RQ3 and RQ4, respectively.

Team C decided to work on 11 of the 13 bugs they had identified in Phase 1. (They did not work on Bug5 or Bug9, so we omit them from our analysis here.) The next subsections present Team C's SES-inclusivity bug fixes, facet by facet, and how well each worked for the Phase 4 participants. Since Team C did not use the "Access" facet to design any of their fixes, we begin with the Communications facet.

#### 5.3.1 Communication Literacy/Education/Culture (Comm) facet: The good and the ugly

Facets drove Team C's fixes in Phase 3; they based their redesigns on the facets they had associated with that inclusivity bug during Phase 1. The bugs that the Phase 1 evaluation had associated with the Comm facet were Bug3, Bug4, Bug10, Bug12, and Bug13. Most of these bugs also involved other facets besides Comm, but since Comm was the only facet for Bug10, we use it as an example, to show a concrete set of fixes Team C used to address Comm-related inclusivity problems.

Bug10's context was in Scenario 2 (Table 2): "*After a <task> submission, student wants to check their submission status and what score they got.*" In Phase 1 using the Original prototype (Figure 3 (left)), the team had identified inclusivity Bug10 using Dav's Comm facet value:

Phase 1 evaluation form: *"He would not be able to identify his grade...jargon. Which column is the actual grade? (Total vs Output Validation)."*

Informed by this evaluation comment, Team C made several communications-oriented changes. They relabeled the columns and tabs to remove jargon; they entirely removed the confusing Output Validation column; they changed vocabulary to the Task names Dav had previously used during submissions; and they used vocabulary such as "Part" and "Task" to clarify relationships between different pieces of information. Figure 3 shows these changes, with the superimposed red boxes pointing them out on the screens.

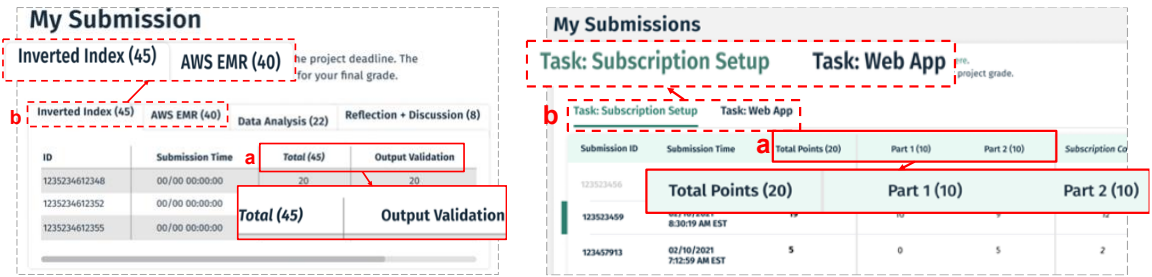

Figure 3: Bug10, Comm facet. (Left): Original; (Right): post-SESMag. Team C changed the column headers (a, red boxes) to remove the confusing "Output Validation" and added "Part #" labeling about the Total Points explanation. They also relabeled tabs (b, dashed red boxes) to remove jargon and to clarify that each tab is about one graded Task. This fix was reasonably successful (see text).

At first glance, the Phase 4 results suggests that these fixes did not help much, because all three of the think-aloud participants who had Dav's Comm facet value—P2, P3, and P6 (as per their questionnaire responses)—ran into trouble on this screen. However, their problems were not with the fixes in Figure 3; nobody ran into problems with those. Instead, they ran into problems that had not been detected in the Phase 1 evaluations. P6 never managed to get to the screen at all because of problems with the menu structure that leads to this screen. P2 and P3 both ran into a problem with mysterious IDs, which Team C had not caught in Phase 1:

P2: *"I wish there was a better ID ... I think this is my submission score but no, I am not quite sure."*

P3: *"... the total points are out of 20, I have 19. But I don't know what <submission ID> one was."*

Thus, the bottom line for Bug10 is that the fixes shown in Figure 3 actually worked for everyone with the Dav Comm facet value, but other problems not detected in Phase 1 arose to trip them up. As to the other bugs related to the Comm facet, the fix for Bug4 turned out to be a resounding success, and Bug3's and Bug12's fixes also fared well. However, Bug13's fixes were a dismal failure.

A Comm problem with Bug13 was vocabulary: the difference between a "primer" (which is not graded) vs. a "project" (which is graded) had not been clear as per Phase 1. Despite the Phase 3 attempts to address that problem, the Phase 4 participants remained confused about the differences and when to work on primers vs. projects. Bug13 later interacted with our assessment of other bugs' fixes, in that it was so troublesome, it stopped some participants from proceeding to steps that could have allowed them to benefit from those fixes.

Table 7 summarizes the results for all bugs tied to the Comm facet.

Table 7: For Comm-related bugs, how well each of Team C's inclusivity bug fixes worked, as measured by number of Phase 4 participants who encountered them in the post-SESMag prototype; denominators reflect number of participants who got to those points. *=see text. Highlighted=bug eradicated for everyone.

| | Fixed for (total participants)... | Fixed for (participants with Dav-Comm facet) |
|---|---|---|
| Bug3 | 6/7 participants (100%) | 2/3 Dav-Comm participants (100%) |
| Bug4 | 6/6 participants (100%) | 3/3 Dav-Comm participants (100%) |
| Bug10 | 3/7 participants (43%) | 0/3 Dav-Comm participants (0%)* |
| Bug12 | 5/6 participants (83%) | 2/2 Dav-Comm participants (100%) |
| Bug13 | 0/7 participants (0%) | 0/3 Dav-Comm participants (0%) |
| Total "tests passed" | 21/33 potential bug instances (64%) | 8/14 Dav-Comm potential instances (57%) |

### 5.3.2 Attitude towards Technology Risks: Inclusivity vs. trade-offs

The Risk facet had arisen often in Phase 1; Team C had used it to identify 7 inclusivity bugs: Bug1, Bug3, Bug4, Bug6, Bug7, Bug8, and Bug13. In designing fixes around the Risk facet, Team C focused on any type of risk that someone like Dav might foresee (except privacy/security risks, which are covered separately by the Priv facet). Most of the fixes focused on the perceived risk of wasting time that Dav does not have to spare, but still not being able to succeed at the task.

Bug3 provides an example of this kind of fix. The context was Scenario 1 (Table 2): "*Student wants to start the first assignment due this week.*" In the Original prototype (Figure 4), the page led off with a boxplot showing how much time (hours) previous students had spent doing this assignment. The boxplot was daunting, suggesting that the project had required up to 24 hours of work by some students. As Team C had mildly put it during Phase 1:

Phase 1 evaluation form: *"might be worried by the hours spent…"*

Bug3 had other aspects that we do not consider here, and other facets were also involved, including Self-Efficacy and Communications/Education/Culture. That said, Team C's fix to the problem shown in Figure 4 did not need to consider other facets: they simply remove the troublesome boxplot.

This fix raises a larger inclusivity point. Although Dav-like students might indeed have their Risk facet value triggered, other students might value a time-estimate feature, to enable them to plan. When aiming for SES-inclusivity, the goal is not to build a system for only Dav-like students, but rather to build a system that works for an entire spectrum of students, from the Dav-like to the Fee-like (recall that Fee is the high-SES counterpart of Dav). This fix may have instead "traded off" these two ends of the spectrum against each other—supporting Dav-like users at the expense of Fee-like users. Ideally, after deciding upon the fix, the team would have used SESMag to evaluate the post-SESMag prototype from *both* Dav's and Fee's perspectives, to enable trade-offs like this should surface. This extra step may have revealed the need for a more nuanced fix than the current one. We will return to this point in the Discussion section.

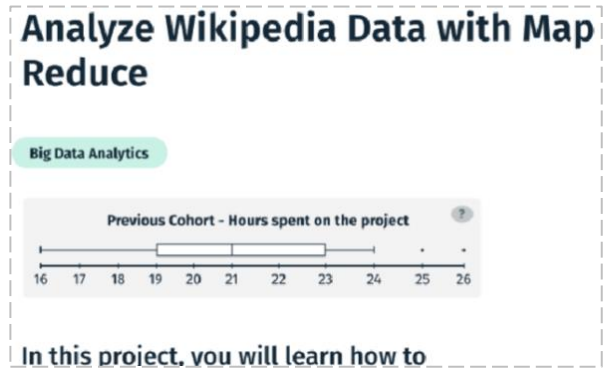

Figure 4: Bug3 and the Risk facet. Original prototype, showing time estimate, which was part of Bug3. A Risk-oriented issue in the Original prototype was this time boxplot feature at the top of each assignment, because it was daunting in both format and content. Team C's fix was simply to remove that feature. This part of the Bug3 fix was successful.

Table 8 shows the results for all seven risk-related bugs. Five of the fixes did very well, many working well for all the participants—especially those with Dav-like Risk facet values. As for the two most troublesome bugs, Bug8 will be discussed in a later section, and we have already discussed Bug13.

Table 8: For Risk-related bugs, how well each of Team C's inclusivity bug fixes worked, as measured by number of Phase 4 participants who encountered them in the post-SESMag prototype. *=see text. Highlighted=bug eradicated for everyone.

|  | Fixed for (total participants)... | Fixed for (participants with Dav-Risk facet)... |
| --- | --- | --- |
| Bug1 | 6/7 participants (86%) | 3/3 Dav-Risk participants (100%) |
| Bug3 | 6/7 participants (86%) | 3/3 Dav-Risk participants (100%) * |
| Bug4 | 6/6 participants (100%) | 2/2 Dav-Risk participants (100%) |
| Bug6 | 6/6 participants (100%) | 2/2 Dav-Risk participants (100%) |
| Bug7 | 6/6 participants (100%) | 2/2 Dav-Risk participants (100%) |
| Bug8 | 3/7 participants (43%) | 1/3 Dav-Risk participants (33%) |
| Bug13 | 0/7 participants (0%) | 0/3 Dav-Risk participants (0%) |
| Total "tests passed" | 33/46 potential bug instances (72%) | 13/18 Dav-Risk potential instances (72%) |

### 5.3.3   Privacy/Security (Priv) -and- Perceived Control/Authority (Ctrl/Auth): One facet value can magnify another

Recall that, informed by Team C's activities during Phase 1, we split off Privacy/Security from other forms of Risk during Phase 1. Using this new facet, Team C found two inclusivity bugs involving the Priv facet: Bug7 and Bug8. Bug8 was particularly interesting, because it also strongly related to Risks aside from Privacy/Security, and to Dav's Perceived Control and Attitudes toward Authority (Ctrl/Auth). This section considers the fixes to Bug8 from all of these perspectives.

For Bug8, the scenario was the same as for Section 5.3.1: "*After a <task> submission, student wants to check their submission status and what score they got.*" In their Phase 1 evaluation, when Team C got to where a student had just submitted the assignment, they then evaluated this subgoal: "*go to the projects grades tab <to check their submission status and score>*". Figure 5 shows the way the screen looked in the Original prototype; note that no option is listed for checking submission status.

The Phase 1 evaluation surfaced two related issues, each tied to a different facet. The first issue was that submitting the file produced no confirmation that the file had been received, so a Dav-like student might wonder if their assignment had even been received. Phase 1 surfaced this problem with the Risk facet (did Dav just waste their scarce time doing the submission wrong? Do they need to waste more time figuring out if they did the right thing?). The second issue was that, if the file had *just* been submitted, how could checking their grade at this point make any sense? The group tied this second issue to Dav's Priv facet value:

R-5 (during Phase 1 evaluation session): "*I'll add a privacy/security <facet> just because ... Dav just submitted this assignment and there's nothing that confirms... not clear for Dav <that checking grade this soon should be an option>*"

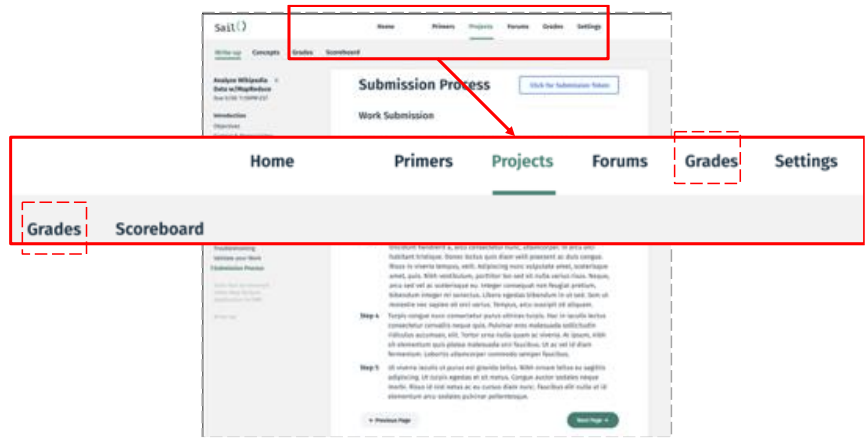

Figure 5: Bug8 and the Priv facet. Original prototype with Bug8 showing, with callouts for magnification. The problem, related to the Priv facet, was that checking submission status was hidden in the "grades" features.

The developers had envisioned Dav going to Grades to check their submission, but Dav does not know this, and may suspect that a Grades option this early (instant grades?) is too good to be true. Although privacy/security concerns are common among all socioeconomic strata, many lower-SES subpopulations are particularly wary, because privacy/security issues arise disproportionately often for them, such as having had their accounts taken over by someone else [7, 41]. The tie to Risk described above could further contribute to Dav's suspicions of something seeming "off".

The Ctrl/Auth facet also played into their analysis—the page had a "Next Page" button that looks likely to be the logical next step, i.e., checking submission status. Dav-like students' attitude toward authority figures, including computer systems as authority proxies, is to follow the authority's directions, since their life experiences have shown that pushing back or deviating from authority's expectations rarely ends well [60, 70]. This attitude could also exacerbate Dav's feelings of Risk in clicking on Grades instead of just following the apparent directions.

Phase 1 evaluation form: *"<Dav expects> The platform will guide Dav ... Assumes it <submission status and score> may be on the next page."*

Team C's fixes in Figure 6 attempted to fix all of the Bug8 issues relating to Risk, Priv, and Ctrl/Auth. They added an explicit way to check status (new Step 6; now operates the same as "Next"). They also relabeled the Grades tab to make clear that it led to the *Project* grade, which covers a group of task submissions, not just this one.

Of the three Phase 4 participants with Dav-like Priv facet values, P7 was entirely successful, and did not run into Bug8. P3 likewise did not run into problems with the Bug8 fixes described here but did run into other parts of the bug not described here. P6, however, was so puzzled with the vocabulary of the "Project Grade" vs "Class Scoreboard" links, that they could not proceed, suggesting that more cleanup of the vocabulary around different kinds of grades is still needed.

P6: *"Is this the part of the project? Then why are there two grades <"Project Grade"> score<Class Scoreboard>?"*

Several other participants also ran into problems with Bug8, with only three of the seven participants having problem-free experiences on this part. In contrast, Team C's fixes to Bug7 (a long series of clicks needed to get to project content) fared much better—no participant experienced any problems with these fixes. The Ctrl/Auth facet was involved in Bug3, Bug6, Bug8, and Bug12. The fixes to Bug3 and Bug6 succeeded for all participants, and Bug12 was also fairly successful. Table 9 summarizes the results for the Priv facet, and Table 10 summarizes the results for the Ctrl/Auth facet.

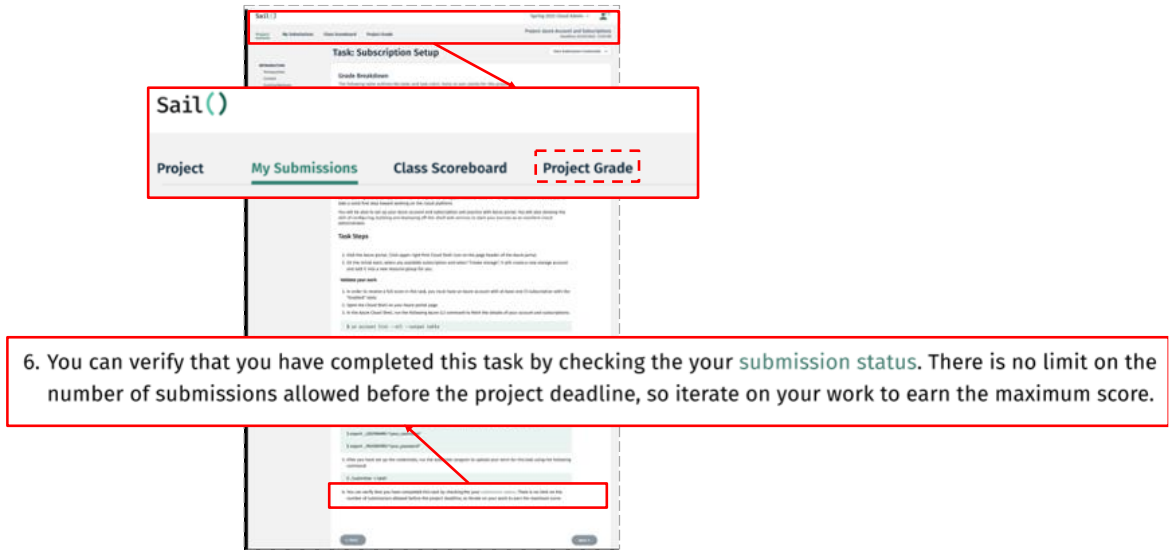

Figure 6: Post-SESMag prototype showing fixes to Bug8, with callouts for magnification. For Privacy, Risk, and Ctrl/Auth, the post-SESMag version of the Bug8 screens conceptually separated grades from submissions. It also cleaned up the two undifferentiated "Grades" links from before (which had oddly led to different places), by clarifying that the *Project* grade (which has several task submissions, not just this one) was check-able. Clicking on Step 6 at bottom leads to the Submissions screen that was shown in Figure 3(right), which lists both pending (ungraded) submissions and graded ones. These fixes made progress, but problems remained despite them.

Table 9: For Privacy/Security-related inclusivity bugs, how well each of Team C's fixes worked, as measured by number of Phase 4 participants who encountered them in the post-SESMag prototype. *=see text. Highlighted=bug eradicated for everyone.

| | Fixed for (total participants)... | Fixed for (participants with Dav-Priv facet)... |
|---|---|---|
| Bug7 | 6/6 participants (100%) | 2/2 Dav-Priv participants (100%) |
| Bug8 | 3/7 participants (43%) | 1/3 Dav-Priv participants (66%)* |
| Total "tests passed" | 9/13 potential bug instances (69%) | 3/5 Dav-Priv potential instances (60%) |

Table 10: For Ctrl/Auth-related inclusivity bugs, how well each of Team C's fixes worked, as measured by number of Phase 4 participants who encountered them in the post-SESMag prototype. *=see text. Highlighted=bug eradicated for everyone

| | Fixed for (total participants)... | Fixed for (participants with Dav-Ctrl/Auth facet)... |
|---|---|---|
| Bug3 | 6/6 participants (100%) | 2/2 Dav-Ctrl/Auth participants (100%) |
| Bug6 | 6/6 participants (100%) | 2/2 Dav-Ctrl/Auth participants (100%) |
| Bug8 | 3/7 participants (43%) | 2/3 Dav-Ctrl/Auth participants (66%)* |
| Bug12 | 5/6 participants (83%) | 2/3 Dav-Ctrl/Auth participants (66%) |
| Total "tests passed" | 20/25 potential bug instances (80%) | 8/10 Dav-Ctrl/Auth potential instances (80%) |

### 5.3.4 Technology Self Efficacy (SE): A string of successes

Technology self-efficacy—a person's beliefs about how good they will be at technology tasks like the one at hand—can serve as a self-fulfilling prophecy. Research has shown that low-SES individuals' life experiences may result in low technology self-efficacy [41, 42], and our instructor/student interviews likewise described having seen/experienced 12 different instances of it among SES-diverse students.

During Phase 1, Team C had unearthed seven inclusivity bugs using of the SE facet: Bug1, Bug2, Bug3, Bug4, Bug6, Bug7, and Bug11. We focus here on Bug11, whose only identifying facet was SE.

The context for Bug11 was in Scenario 3: *"Halfway into the semester, student wants to check if they will pass this course"*. In the Original version of the prototype, the Grades screen displayed all the projects and the corresponding grades for the student, and Team C identified checking their grade on this page as a first subgoal. However, they anticipated that:

Phase 1 evaluation form: *"They might want to check their grade but also feel not confident on checking it out. (And instead want to ask the instructor)"*

The team also noted that once Dav arrived at the page, there was no scale/indication on the grades screen to measure "passing grades".

Team C made several changes to their prototype in response to these Phase 1 results. First, they renamed the "Grades" to "Course Grade" to avoid confusion with project grades. Second, they also added progress bars along with colors (green vs. red) for each project, to indicate whether the student had passed or failed the project. By signaling success/failure more explicitly, Team C aimed to increase Dav-like users' confidence that they understood their status in the course (Figure 7).

Team C's design changes worked well for Dav-like participants. In the Phase 4 study, none of the participants with Dav-like SE facet values (P2, P4, or P5), ran into any of Bug11's problems. In fact, Team C's SE-related fixes worked for all participants with Dav's SE facet value in all six SE-related bugs—and for all except two of the 45 possible encounters for *any* participant (Table 11).

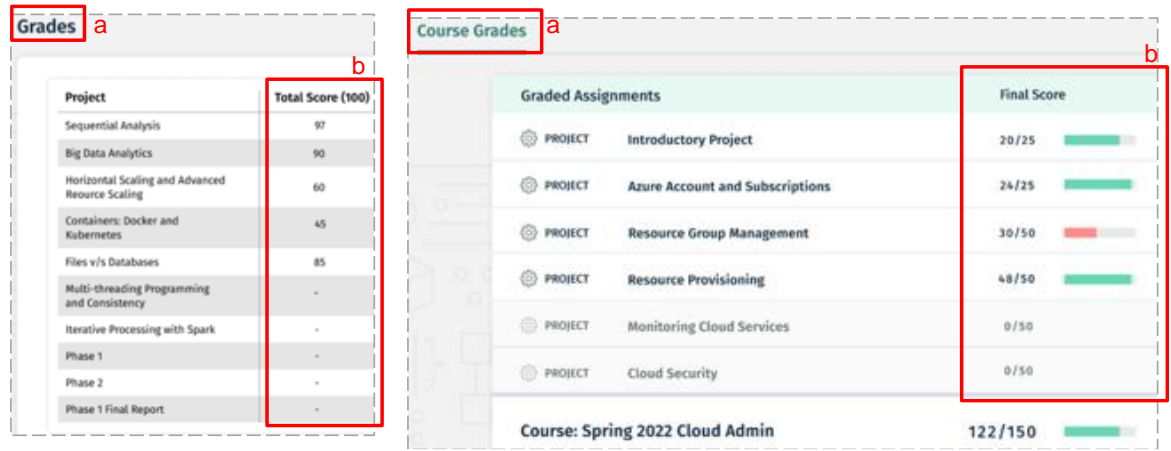

Figure 7: Bug11 and the SE facet. (Left) In the Original version of the prototype (a) under Grades, all projects were listed along with (b) scores. (Right) In the post-SESMag version of the prototype, Team C (a) renamed grades to "Course Grades" and (b) added colored progress bars for each project (green vs red) to aid Dav like users to understand their progress in the course. This fix was very successful, fixing the bug for everyone who could have encountered it.

Table 11: For SE-related inclusivity bugs, how well each of Team C's fixes worked, as measured by number of Phase 4 participants who encountered them in the post-SESMag prototype. *=see text. Highlighted=bug eradicated for everyone

|  | Fixed for (total participants)... | Fixed for (participants with Dav-SE facet)... |
|---|---|---|
| Bug1 | 6/7 participants (86%) | 3/3 Dav-SE participants (100%) |
| Bug2 | 7/7 participants (100%) | 3/3 Dav-SE participants (100%) |
| Bug3 | 6/7 participants (86%) | 3/3 Dav-SE participants (100%) |
| Bug4 | 6/6 participants (100%) | 3/3 Dav-SE participants (100%) |
| Bug6 | 6/6 participants (100%) | 3/3 Dav-SE participants (100%) |
| Bug7 | 6/6 participants (100%) | 3/3 Dav-SE participants (100%) |
| Bug11 | 6/6 participants (100%) | 3/3 Dav-SE participants (100%) |
| Total "tests passed" | 43/45 potential bug instances (96%) | 21/21 Dav-SE potential instances (100%) |

# 6 DISCUSSION

Team C identified 13 SES-inclusivity bugs (RQ1), validated them (RQ2), and attempted to fix (RQ3) 11 of them. Tables 8–11 (or Appendix A's Table A-2) show that the bugs these fixes targeted entirely disappeared for all participants (5/11, or 45%) (RQ4). Further, 6/11 (54%) were successful for same-facet participants (RQ4). What RQ1 and RQ2 say is that SESMag was quite effective in revealing SES-inclusivity bugs. What RQ3 and RQ4 say is that SESMag had mixed success in carrying Team C through to fixing them. We reflect upon these points next, and then discuss SESMag's inclusivity strategy and limitations of our study.

## 6.1 The SESMag method's accuracy: soundness and completeness

We consider SESMag's accuracy from two perspectives: soundness and completeness. For this case study, soundness from the perspective of the bugs SESMag reports, means that, if Phase 1 identified an inclusivity bug, that inclusivity bug will really happen to some user. For example, a reader might wonder if the team's SESMag reasoning (e.g., in Section 5.1's Table 6) would really arise with actual users. If not, that would be a false positive. The more false positives, the less soundness.

Fortunately, because the SESMag process is built upon cognitive walkthroughs, there is reason to believe that SESMag will inherit cognitive walkthroughs' very low false-positive rate. Research shows that humans performing methods in the CW- family report few false positives (10% or below across numerous studies) [49]. This low false-positive rate has also been seen in GenderMag, a sibling of SESMag, whose false positive rate to date has run 5% or less [11, 86]. The Phase 2 instructor/student's experiences (Section 5.2) add further evidence as to the soundness of the Phase 1 results. On the other hand, soundness from other perspectives remain open questions. For example, our case study did not have enough participants to analyze whether the bugs found actually do occur disproportionately often in low-SES users, or whether the facets the evaluators used to identify the bugs would actually be the facets that arose in a participant when they encountered a bug. We leave these questions for future research.

Turning to completeness from the perspective of the inclusivity bugs found, lack of completeness would mean that evaluators in Phase 1 overlooked an inclusivity bug that actually arose later among the users—a false negative. We pointed this out with Bug10 in Section 5.3.1, but did not analyze this further in our case study. However, it is not unexpected, since research has shown that other members of the cognitive walkthrough can suffer from as high a false negative rate as 70% [49]. Thus, we have little cause for optimism as to SESMag's completeness, i.e., its ability to reveal *all* the SES-inclusivity bugs a product has. Still, we argue that for finding inclusivity bugs, soundness is more important than completeness. If using the SESMag method early enough to fix SES-inclusivity bugs found removes even a handful of inclusivity bugs, the experiences for SES-diverse users will be better than they otherwise would have been.

## 6.2 Reflecting on "Facets drive fixes"

The SESMag method generates, for each detected inclusivity bugs, a facet value can explain *why* a person might be excluded by the bug, which can point toward a fix direction. Team C harnessed these facet values, using the facet values that had identified the inclusivity bug to "drive" its fix for that bug. Team C's pattern of successes and failures in fixing the bugs suggests that this process is easier for some facet values than others.

Tables 8–11 together (brought together for convenience in Appendix A's Table A-2) show that Team C's fixes were very successful for participants with Dav-like Access, Ctrl/Auth, and SE facet values—in the case of the SE facet, remarkably well (recall Table 11). However, participants with Dav-like Comm, Risk, and Priv facet values did not fare as well, perhaps because some facet values may be (1) inherently more difficult to accommodate for, (2) more difficult for designers to understand, or (3) conflict with other facets and force tradeoff decisions. We touch on the latter hypothesis again in the next subsection.

## 6.3 SESMag's Inclusivity Strategy—the Personas as Endpoints

SESMag personas are different from most other personas. Unlike personas designed to be representative of some group of users, SESMag personas are instead designed to be particular collections of "opposite" facet values, one at each endpoint of a wide range of values for that facet. For example, Dav is risk-averse, and Fee is risk-tolerant.

SESMag, like other members of the InclusiveMag family, aims to help practitioners create products that serve wide ranges of users' attitudes and preferences (e.g., their attitudes toward risk). The method allows for consideration of both endpoints of an ordinal scale (e.g., from low risk aversion to high risk aversion), by using two personas with opposite values for every facet. The idea is that any feature that *simultaneously* works for both endpoints of the facet value range will work for the facet values between those endpoints. For example, if practitioners can successfully iterate on a design until a feature simultaneously supports both those who enjoy risk and those most who strongly avoid it, then it likely supports everyone's level of risk-aversion.

Thus, although each of Dav's and Fee's facet values are evidence-based syntheses of lower-SES and higher-SES users, respectively, neither Dav nor Fee is intended to be "representative" as personas. Most real users will not have the full complement of either Dav's or Fee's extreme facet values, but will likely fall somewhere between Dav and Fee, and/or have a mix of Dav's and Fee's facet values. For example, someone might be tolerant of risk, like Fee, but have little access to reliable technology, like Dav.

However, in our study, Team C used only the Dav persona, to embody the low-SES end of every facet's value range. For Team C, the Fee-like counterpart was their own prior experiences supporting higher-SES students. This kind of short-cut saved Team C evaluation time, but may have come at a cost: overfocusing on Dav at the expense of Fee. Section 5.3.2 showed a case in point.

For teams who do use both personas, is the utopian vision of universal inclusivity for every value of every facet always possible? Research with SESMag's sibling, GenderMag, has shown that software can often be made considerably more inclusive and equitable using this approach [86], but there is no a priori reason to expect that *every* inclusivity bug can be solved with a redesign in a way that does not sacrifice inclusivity in another way. Returning to the example in Section 5.3.2, Team C's Original prototype's time estimates for projects could discourage a person daunted by the size of these estimates, but encourage someone who prefers information like this to plan ahead. Maybe a novel design could satisfy both; maybe not. But if universal inclusivity is not possible, a facet-based analysis can at least let a design team make informed decisions about what kinds of people their prototype is supporting.

### 6.4 Limitations

No empirical study is perfect. One reason is the inherent trade-off among the kinds of findings different types of studies can produce. Field studies aim for real-world applicability, but at the cost of not being able to isolate different variables. Controlled lab studies, on the other hand, can isolate controls but cannot measure real-world applicability. Our field study therefore had uncontrolled variables, mainly in the form of Team C's real-world timing and availability constraints.

Evidence for improvement of the software would have been stronger if Phase 4 had been an "Original vs. Post-SESMag" between-subjects study, because that would have given a more direct comparison of improvement between the Original prototype's inclusivity bugs vs. those of the Post-SESMag prototype. However, we had difficulty attracting enough participants to do this. We mitigated this by validating that the Original version's bugs (found in Phase 1) were real, using the Phase 2 interviews with an instructor and a community college student. This allowed us to know that the bugs existed, and we then measured in Phase 4 whether those bugs still existed. Still, it is possible that Phase 1 missed some class of bugs that a pre-change thinkaloud study might have caught.

A possibility of biasing the instructor's or student's responses to our interview questions also arises in Phase 2. These questions came directly from the output of Phase 1. The questions were asked to provide enough context to the interviewees to understand the problem. However, this context may have led to providing extra information that might have biased the opinion of the interviewees. The interview study may also have biased interviewees towards confirming problems suggested by the walkthrough. To mitigate these risks, we instructed the interviewees to report historical information (what they had seen or experienced). The fact that interviewees' responses often included personal anecdotes and informally referenced the facets, which we did not share with them, provided further evidence that interviewees were not simply being agreeable.

Participant recruitment was subject to selection bias: students willing to participate in such a study may have different characteristics than students who would not participate in such a study.

Field studies like ours also have limitations in generalizability. Our study included a particular team (Team C), and a particular platform (the learning platform), and aimed at a particular population (SES-diverse community college students). Thus, interpretations we made from our data might be different had we worked with a different team, used a different platform, or aimed at a different population. Given these limitations, we do not view our results as being generalizable beyond the particular context of our investigation, but rather as encouraging evidence of the SESMag approach's promise.

### 7 CONCLUSION

In this paper, we contribute the first evidence that the SESMag method can be used to find real inclusivity bugs that affect low-SES users, and provides sufficient information to guide a designer towards fixing these bugs.

Specifically, Phase 1 of our case study of Team C's online learning management platform showed that (RQ1:) the practitioners were indeed able to use SESMag to identify SES-inclusivity bugs disproportionately faced by low-SES individuals. In fact, Team C found SES-inclusivity bugs in 76% of the software features they evaluated. Phase 2 showed that (RQ2): these bugs were real: interviews with a college instructor and with a low-SES college student revealed that one or both of them had actually experienced or had witnessed the 13 inclusivity bugs Team C had found. Phase 3 revealed that (RQ3): Team C was able to follow the SES facets at the SESMag method's core to derive inclusivity fixes. The user study with SES-diverse users in Phase 4 showed dramatic reductions in the number of inclusivity bugs a group of SES-diverse community college students faced after the fixes had been made. 45-54% of the inclusivity bugs completely disappeared (Appendix A's Table A-2). From a facet-instance perspective, these tables show that the success rate ranged from 64% (Comm facet) to 96% (SE facet).

Perhaps most important, this investigation shows that many SES-inclusivity bugs take the form of cognitive or access barriers that the software arbitrarily imposes, such as requirements that all users have a certain amount of risk tolerance, a certain amount of reliability in their devices/wifi, a certain comfort level with privacy/security risks, a certain level of confidence, etc. Our results suggest that software's SES-inclusivity can improve by finding such inclusivity bugs and removing them—in an inclusivity-debugging approach.

Still, as the first field study on the emerging SESMag method, this paper is only a first step toward investigating to what extent SESMag can make technology more equitable for SES-diverse users. We invite other researchers and practitioners to join us in carrying out further investigations into this question.

## ACKNOWLEDGMENTS

Anonymized

# APPENDIX A

Table A-1 contains the complete version of the main paper's Table 6, which was abbreviated to conserve space.

Table A-1: SES-inclusivity bugs, facets, and triangulation. Columns 1-2: Phase 1 results. Column 3: triangulation with research literature. Column 4: Phase 2 data. Note Phase 1 results' consistency with research and with problems witnessed/experienced by the Instructor (I) and/or Student (S). (This is an expanded version of the main paper's Table 6.)

| Phase 1 data: | | Consistency with: | |
|---|---|---|---|
| **SES-inclusivity bugs:**<br>**Someone like Dav...** | **Facet values (Table 3 abbreviations)**<br>**+ Why (excerpted from evaluation forms)** | **Research Literature**<br>**relating to excerpt** | **I/S** |
| *Bug1*: Might not start the first assignment due this week: Big Data Analytics. | Risk: "Dav doesn't have <that much> time… grocery store shift starts in an hour…"<br>Other facets used: SE | Risk includes aversion to <u>wasting time</u> on tech dead-ends [3, 6, 15] | I |
| *Bug2*: Might not click the "ongoing" button for Big Data Analytics. | Access: "...using an <u>ancient laptop</u>… flaky keyboard... may actually be pounding on <unclickable> Big Data Analytics"<br>Other facets used: SE | Access: seldom access to <u>reliable device</u>(s)/software & internet [34] | I S |
| *Bug3*: After clicking "ongoing" button, might not know was right/see progress. | SE: "I don't see the word "assignments", so <u>I'd be doubtful</u> this was right."<br>Other facets used: Risk, Comm, Ctrl/Auth | SE: lower tech self-efficacy including <u>underestimating ability</u> to succeed [42] | I S |
| *Bug4*: Might not read through the page and click Next. | Comm: "maybe <u>scared by techy words</u> he doesn't know: MapReduce, Parallel Processing. Stressful vocab"<br>Other facets used: SE, Risk | Comm: understanding the <u>wording</u> (including jargon) & education [27, 43] | I S |
| *Bug5*: After reading/clicking Next, not know was right/see progress. | Comm: "<u>will not read "why"</u> ... huge wall of text"<br>Other facets used: Risk | Comm: see Bug4 [27, 43] | I S |
| *Bug6*: Might not scroll down, read prerequisites, and click on MongoDB [primer] | Risk: "Will scroll down, read prereqs, but not click on MongoDB; will start project first, <u>due to limited time</u>."<br>Other facets used: SE, Ctrl/Auth | Risk: see Bug1 [3, 6, 15] | I S |
| *Bug7*: After scrolling, reading, clicking MongoDB, might not know was right/see progress. | Risk: "<u>Hitting "next" on and on</u> seems like the scenic route; wishes he could click the task directly."<br>Other facets used: SE, Privacy | Risk: see Bug1 [3, 6, 15] | I S |
| *Bug8*: Might not check ... after a project submission... what score they got. | Privacy: "It is not clear for dav <that grades tab makes sense here>... no indication that the file was uploaded/ received, so…a little confused <u>risk to privacy and security."</u><br>Other facets used: Risk, Ctrl/Auth | Privacy: includes <u>distrust of losing personal info</u> to unexpected features [33,60] | I S |
| *Bug9*: Might not click "grades" near the upper left corner. | Ctrl/Auth: "There's <u>no indication</u> that he made a submission..."<br>Other facets used: Risk | Ctrl/Auth: includes <u>low perception of control over tech outcomes</u> [33, 60] | I S |
| *Bug10*: Might not read submissions table & identify grade. | Comm: "He would read <but> not be able to identify his grade, might <u>not understand all the tech jargon</u>"<br>Other facets used: — | Comm: see Bug4 [27, 43] | I S |
| *Bug11*: Might not check halfway into term if they'll pass | SE: "might want to check their grade but also <u>don't feel confident on <how</u>, so> instead want to ask ...instructor"<br> | SE: see Bug3 [42] | S |
| *Bug12*: After clicking "grades" in navigation, might not know was right/see progress. | Ctrl/Auth: "There is <u>no indication</u> whether these are <u>considered 'passing score'</u>..."<br>Other facets used: Comm | Ctrl/Auth: see Bug9 [33,60] | I S |
| *Bug13*: After scanning table for assignments due, might not know was right/see progress. | Comm: "…not sure whether he needs to do <u>projects vs primers</u> and which assignment type is going to be graded."<br>Other facets used: Risk | Comm: see Bug4 [27, 43] | I S |

Table A-2 brings Tables 8–11 together, all in one place. Each row represents a bug, and shows the facets involved, the people in Phase 4 for whom the bug was resolved (broken down by who did or did not have low-SES facets matching the bug's facet).

For example, the first row shows that Team C's fix for Bug1, involving Risk and SE facets, successfully prevented any difficulties (column 3) *any* participant with Dav-like Risk or SE facet values, and for two other participants (Fee-like for those facet values) as well. However, one Fee-like participant (column 4) did run into difficulties with that bug.

Table A-2: How well each of Team C's inclusivity bug fixes from Phase 3 worked, for Phase 4's participants. "Now no problems" includes Table 4's codes √ and√?. (Fixed are out of fewer than 7 participants if someone did not take a path where they could experience it.)

| | Bug for... (Phase 1) | Fix result (Phase 4): Now no problems for participants... | Fix result (Phase 4): Still problem and/or new problems introduced for participants... | Fixed for... All | Facet |
|---|---|---|---|---|---|
| Bug1 | Risk, SE | ...with these relevant Dav facets: P2-_RiskSE_, P4-_RiskSE_, P5-_SE_, P7-_Risk_ ...without them: P1, P6 | ...without these Dav facets: P3 | 6/7 (86%) | 4/4 (100%) |
| Bug2 | Access, SE | ...with these relevant Dav facets: P1-_Acc_, P2-_AccSE_, P4-_SE_, P5-_SE_, P7-_Acc_ ...without them: P3, P6 | - | 7/7 (100%) | 5/5 (100%) |
| Bug3 | Access, Comm, Risk, Ctrl/Auth, SE | ...with these relevant Dav facets: P1-_Acc_, P2-_AccCommRiskCtrlSE_, P3-_Comm_, P4-_RiskSE_, P5-_CtrlSE_, P7-_AccRisk_ | ...with these relevant Dav facets: P6-_Comm_ | 6/7 (86%) | 6/7 (86%) |
| Bug4 | Comm, Risk, SE | ...with these relevant Dav facets: P2-_CommRiskSE_, P3-_Comm_, P4-_RiskSE_, P5-_SE_, P6-_Comm_ ...without them: P1 | - | 6/6 (100%) | 4/4 (100%) |
| Bug6 | Risk, Ctrl/Auth, SE | ...with these relevant Dav facets: P2-_RiskCtrlSE_, P4-_RiskSE_, P5-_CtrlSE_ ...without them: P1, P3, P6 | - | 6/6 (100%) | 3/3 (100%) |
| Bug7 | Risk, Priv, SE | ...with these relevant Dav facets: P2-_RiskSE_, P3-_Priv_, P4-_RiskSE_, P5-_SE_, P6-_Priv_ ...without them: P1 | - | 6/6 (100%) | 5/5 (100%) |
| Bug8 | Risk, Priv, Ctrl/Auth | ...with these relevant Dav facets: P5-_Ctrl_, P7-_RiskPrivCtrl_ ...without them: P1 | ...with these relevant Dav facets: P2-_RiskCtrl_, P4-_Risk_, P3-_Priv_, P6-_Priv_ | 3/7 (43%) | 3/6 (50%) |
| Bug10 | Comm | ...without these Dav facets: P1, P5, P7 | ...with these relevant Dav facets: P2-_Comm_, P3-_Comm_, P6-_Comm_ ...without them: P4 | 3/7 (43%) | 0/3 (0%) |
| Bug11 | SE | ...with these Dav facets: P2-_SE_, P4-_SE_, P5-_SE_ ...without them: P1, P3, P7 | - | 6/6 (100%) | 3/3 (100%) |
| Bug12 | Comm, Ctrl/Auth | ...with these relevant Dav facets: P3-_Comm_, P5-_Ctrl_, P7-_Ctrl_ ...without them: P1, P4 | ...with these relevant Dav facets: P2-_CommCtrl_ | 5/6 (83%) | 3/4 (75%) |
| Bug13 | Comm, Risk | - | ... with these relevant Dav facets: P2-_CommRisk_, P3-_Comm_, P4-_Risk_, P6-_Comm_, P7-_Risk_ ...without them: P1, P5 | 0/7 (0%) | 0/5 (0%) |
| Total by "fixed-for" | Fixed for 54/72 (75%) of potential bug _instances_; and for 36/49 (73%) with these facets | | | 54/72 | 36/49 |
| Total by bugs gone | Eradicated 5/11 (45%) of inclusivity bugs for all; and 6/11 (54%) for same-facet participants | | | 5/11 | 6/11 |